\documentclass[aps,pre,reprint,10pt,twocolumn,showpacs,nofootinbib]{revtex4-1} 

\usepackage{amsmath}
\usepackage{amsfonts}
\usepackage{amssymb}
\usepackage{graphicx}
\usepackage{subfigure}
\usepackage[pdfborder={0 0 0}]{hyperref}  
\usepackage{color}
\usepackage[latin1]{inputenc}
\usepackage{float}
\usepackage{comment}

\begin{document}

\title{The Hilbert-Schmidt norm as a measure of entanglement in spin-$1/2$ Heisenberg chain: generalized Bell inequality and distance between states}
 
\begin{abstract}
In this letter, we show that the measure of entanglement using the generalized Bell inequality and the distance between states coincide when we use the Hilbert-Schmidt norm. Our conclusions apply to the spin-$1/2$ Heisenberg chains with interaction between the first neighbors. 
\end{abstract}

\author{Saulo Luis Lima da Silva}
\email{saulosilva@cefetmg.br}
\affiliation{Centro Federal de Educação Tecnológica de Minas Gerais,\\
	Avenida Monsenhor Luiz de Gonzaga, 103 -- centro -- 37250-000 --
	Nepomuceno -- MG -- Brazil.}

\author{Daniel H. T. Franco}	
\email{daniel.franco@ufv.br}
\affiliation{Departamento de Física, Grupo de Física-matemática e Teoria Quântica de Campos, Universidade Federal de Viçosa, CEP 36570-900,
	Viçosa -- MG, Brazil}

\maketitle

\section{Introduction}
Since the end of the last century we have seen with great enthusiasm the great achievements in the field of quantum information. Remarkably, much effort was devoted to the understanding, measurement, and control of entanglement \cite{ref1, ref2, ref3, ref4, ref5, ref6, ref7, ref8}. Despite this, we have not yet been able to obtain a unique form of characterization and measure of entanglement. What we have are criteria that, in some cases, allow us to determine if there is entanglement in the system. A major breakthrough was struck by the work of the Horodecki family in 1996 by showing what became known as the Peres-Horodecki criterion \cite{ref2}. They showed that for systems whose dimension in the Hilbert space is not greater than $2\otimes3$, the positivity of the partial transpose of the reduced density matrix of the system is necessary and sufficient condition for the entanglement.
   
The Peres-Horodecki criterion allows us to qualitatively determine entanglement. There are several proposals in the literature for the quantitative determination of entanglement \cite{ref3, ref4, ref5, ref9, ref10}. One of the best known is the formation entanglement, which in the case of two qubits lies in the well-known Wootters formula \cite{ref5}. Although there are generalizations of concurrence for $s>1/2$ \cite{refLi, refOst,refBah}, there are so far no simple expressions \cite{refSch}. A more general proposal is the distance between states as a measure of entanglement \cite{ref3}. In this case, the distance between the state of interest and the set of separable states is used as a measure of the degree of entanglement of the system. This proposal has a strong geometric appeal and the advantage of not being limited to the size of the system or the number of particles of the system. The disadvantage is that numerical methods are usually required for the calculation of the distance between states, since the determination of the separable matrix set element closest to the state of interest is not trivial. 

In recent work, we have shown an analytical way to quantitatively calculate thermal and macroscopic entanglement using the distance between states \cite{ref10,refSau}. For this, we made use of the Peres-Horodecki criterion, which allows us to analyze systems larger than $2\otimes2$. We use the Hilbert-Schmidt norm as a measure of the distance between states \cite{ref4, refDahl}. Later, other works were published using this technique to measure spin chain entanglement of various compounds with and without the application of magnetic field \cite{ref10,ref11,ref12,refSau}.
   
In this paper we aim to substantiate the use of the Hilbert-Schmidt norm as a simpler and more natural measure of the distance between states to quantify the entanglement in spin-$1/2$ Hesenberg chain. For this, we will be based on the Bertlmann-Narnhofer-Thirring theorem which relates the Hilbert-Schmidt norm to the maximum value of the violation of a generalized Bell inequality (GBI)\cite{ref13}.

The outline of this letter is as follows. In Section II we describe the generalized Bell inequalities (GBI). In Section III, we show that the entanglement measured by GBI and the distance between states coincide. Section IV is dedicated to the conclusions.

\section{Generalized Bell Inequalities}

\begin{figure}[ht!]
	\centering
	\includegraphics[scale=0.5]{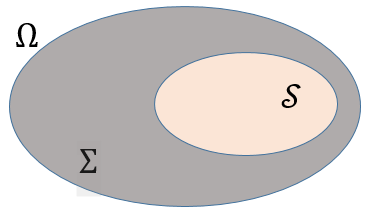} 
	\caption{Illustrative picture of the sets $\Omega$, $\Sigma$ e $\mathcal{S}$.}
	\label{fig1}
\end{figure}

According to Bertlmann-Nernhofer-Thirring \cite{ref13} for the construction of a generalized Bell inequality (GBI) consider a finite dimensioned Hilbert space, $\mathcal{H}=\mathcal{C}^N$, where the observables $A$ are represented by all hermitian matrices and the states $\sigma$ by densities matrices. Let $\Omega$ be the set of density matrices constituted by the set of separable density matrices $\mathcal{S}$ and by the set of entangled density matrices $\Sigma=\Omega-\mathcal{S}$, as shown in Figure \ref*{fig1}. 

The scalar product is defined by $$\langle \rho,A \rangle=\mathrm{Tr}(\rho A),$$ and the corresponding norm is $$\|A\|=\bigl(Tr\,A^2\bigr)^{1/2}.$$
Bell's inequality in a generalized sense is given by an $A \ngeqslant 0$ operator for which
\begin{equation}\label{eq11}
\langle \rho,A \rangle \geqslant 0\,\,,\,\,\forall\,\rho \in {\cal S}.
\end{equation} 

So there is $\sigma \in \Sigma$ such that $$\langle \sigma, A \rangle < 0.$$
So GBI (\ref*{eq11}) can be violated by a tangled state $\sigma \in \Sigma$. Soon we can write the following inequality 
$$\langle \rho,A \rangle > \langle \sigma, A \rangle.$$ 
Considering the maximum violation of GBI 
$$ B(\sigma,\rho)=\max_{\|A-\alpha {\mathbf{1}}\| \leqslant 1}
\Bigl(\min_{\rho \in {\cal S}}\langle \rho,A \rangle - \langle \sigma,A \rangle \Bigr),\,\,\mbox{with $\alpha \in \mathbb{R}$},$$
Bertlmann-Narnhofer-Thirring have proven that \cite{ref13}
\begin{itemize}
		\item The maximum violation of GBI is equal to the distance from $\sigma$ to the set $\mathcal{S}$, ie $$B(\sigma,\rho)=D(\sigma,\rho),\,\,\forall\,\sigma \in \Sigma.$$
	
	    \item The minimum of $D$ is obtained for some $\rho_0$ and the maximum of $B$ for 
	    \begin{equation}\label{eq2}
    	A_{\max}=\frac{\rho_{_0}-\sigma-\langle \rho_{_0},(\rho_{_0}-\sigma) \rangle{\mathbf{1}}}{\|\rho_{_0}-\sigma\|}.
    	\end{equation}
	
	    \item For $D=B$ we have the following 
    	\begin{equation}\label{eq3}
    	\min_{\rho \in {\cal S}} \left\{\rho-\sigma \Bigl| \frac{\rho^\prime-\sigma}{\|\rho^\prime-\sigma\|}\right\}
    	\leqslant B(\sigma,\rho) \leqslant \|\rho^\prime-\sigma\|,\,\,\forall\,\rho^\prime \in {\cal S}.
    	\end{equation}
\end{itemize}

\section{Distance between states, generalized Bell inequalities and Heisenberg chains}

Now let's show that the entanglement in Heisenberg chains, $\mathcal{E}(\sigma)$ measured using the maximum violation of generalized Bell inequality, $\mathcal{B}(\sigma, \rho)$, is equal to the entanglement in Heisenberg chains measured by the distance between states using the Hilbert-Schmidt norm, $\mathcal{D}(\sigma, \rho)$. Thus, we justify the Hilbert-Schmidt norm as a measure for the entanglement of a system.

Consider the spin-$1/2$ antiferromagnetic Heisenberg chain $1-D$ modeled by \begin{equation}
H_{\rm AF} = -J \sum_i \vec{S_i} \cdot \vec{S}_{i+1}~, \label{H}
\end{equation}
where $J<0$ is the exchange coupling constant and $S_i$ is the spin operator for the $i$th spin (on site $i$).

As shown and discussed in \cite{ref10} the set of separable and entangled matrices can be written (making use of the Peres-Horodecki criterion) respectively as

\begin{equation}
\rho =
\begin{pmatrix}
v_{s} & 0 & 0 & 0 \\
0 & w_{s} & z & 0 \\
0 & z^* & w_{s} & 0 \\
0 & 0 & 0 & v_{s}
\end{pmatrix}~,\label{matrixs}
\end{equation}
and
\begin{equation}
\sigma =
\begin{pmatrix}
v_{e} & 0 & 0 & 0 \\
0 & w_{e} & z & 0 \\
0 & z^* & w_{e} & 0 \\
0 & 0 & 0 & v_{e}
\end{pmatrix}\label{matrixe}
\end{equation} 
where $v_{e}<\vert z\vert$ and $v_{s}\geq  \vert z \vert$. Moreover $$v = \frac{1}{4}+\langle S_{i}^{z}S_{j}^{z}\rangle,$$ $$z = 2\langle S_{i}^{x}S_{j}^{x}\rangle$$ and $$w=\frac{1}{2}-v.$$

To quantify the entanglement of a state $\rho_e$ we will use Hilbert-Schmidt norm as a measure of the distance between this state and the set of separable density matrices,

\begin{equation}
\mathcal{D}(\rho_s, \rho_e)=\sqrt{Tr[(\rho_s - \rho_e)^2]}=2 \vert v_s - v_e \vert.
\end{equation}
Entanglement will be given by the minimum of this distance. The minimum occurs when $v_s = \vert z \vert$ \cite{ref10}. Thus

\begin{equation}\label{eq1}
\mathcal{E}(\rho_e) = 
\begin{cases}
2(\vert z \vert - v_e), & v_e < \vert z \vert. \\
0, & v_e \geq \vert z \vert. 
\end{cases}
\end{equation}


For the calculation of the GBI consider

\begin{equation}
\rho_0=
\begin{pmatrix}
\vert z \vert & 0 & 0 & 0 \\
0 & 1/2-\vert z \vert & z & 0 \\
0 & z^* & 1/2-\vert z \vert & 0 \\
0 & 0 & 0 & \vert z \vert
\end{pmatrix}.
\end{equation}
Soon

\begin{equation}
\rho_0-\sigma=
\begin{pmatrix}
\vert z \vert-v_e & 0 & 0 & 0 \\
0 & v_e-\vert z \vert & 0 & 0 \\
0 & 0 & v_e-\vert z \vert & 0 \\
0 & 0 & 0 & \vert z \vert-v_e
\end{pmatrix},
\end{equation}
and

\begin{equation}
\langle\rho_0,(\rho_0,\sigma)\rangle\mathbf{1}=(\vert z \vert -v_e)(4\vert z \vert-1)
\begin{pmatrix}
1 & 0 & 0 & 0 \\
0 & 1 & 0 & 0 \\
0 & 0 & 1 & 0 \\
0 & 0 & 0 & 1
\end{pmatrix},
\end{equation}
and

\begin{equation}
\vert \vert \rho_0 -\sigma \vert \vert=2(\vert z \vert -v_e).
\end{equation}
Therefore

\begin{equation}
A=
\begin{pmatrix}
1-2\vert z \vert & 0 & 0 & 0 \\
0 & -2\vert z \vert & 0 & 0 \\
0 & 0 & -2\vert z \vert & 0 \\
0 & 0 & 0 & 1-2\vert z \vert
\end{pmatrix}.
\end{equation}
This allows us to calculate explicity the GBI using the equation (\ref{eq3}). For this, note that

\begin{equation}
\langle \rho,A \rangle = 2(v_s-\vert z \vert),
\end{equation}
as  $v_s\geq \vert z \vert$, we have
\begin{equation}
\mathrm{min}\langle \rho, A \rangle = 0.
\end{equation}
Furthermore

\begin{equation}
\langle \sigma,A \rangle=-2(\vert z \vert-v_e),
\end{equation}
so

\begin{equation}
B(\sigma, \rho)=2(\vert z \vert-v_e),
\end{equation}
which is the same result obtained in (\ref{eq1}), showing that the distance between states using the Hilbert-Schmidt norm as a measure of entanglement coincides with the generalized Bell inequality (GBI),

\[
{\cal E}(\sigma)=B(\sigma,\rho)= \mathrm{min} D(\sigma,\rho)\,\,,
\]
as we wanted to demonstrate. This result is in agreement with that obtained through the Wootters formula \cite{ref5, Wootters}.

\section{Conclusion}
In this work, we present a result that reinforces the use of the Hilbert-Schmidt norm as a measure of entanglement. We show that using this norm, the generalized Bell inequality and the distance between states coincide. Our treatment has been applied to systems that can be modeled by spin-$1/2$ Heisenberg chains. The result obtained is in perfect agreement with what is obtained when using the Wootters formula.

\label{sec:Conc}


\bibliographystyle{apsrev4-1}
\bibliography{ensembles}

\begin{thebibliography}{99}

\bibitem{ref1} A. Aspect, P. Grangier and G. Roger. ``{\it Experimental realization of Einstein-Podolky-Rosen-Bohm gendankenexperiment: a new violation of Bell's inequalities,}'' 
{\bf Phys. Rev. Lett., 49} (1982).
%
\bibitem{ref2} M. Horodecki and P. Horodecki and R. Horodecki. ``{\it Separability of mixed states: necessary and sufficient conditions.,}'' 
{\bf Phys. Lett. A, 223} (1996).
%
\bibitem{ref3} V. Vedral, M. B. Plenio, M. A. Rippin and P. L. Knight. ``{\it Quantifying entanglement.,}'' 
{\bf Phys. Rev. Lett., 78} (1997).
%
\bibitem{ref4} C. Witte and M. Trucks. ``{\it A new entanglement measure induced by the Hilbert-Schmidt norm,}'' 
{\bf Phys. Lett. A, 257} (1999).
%
\bibitem{ref5} W. K. Wootters. ``{\it Entanglement of formation of an arbitrary state of two qubits.,}'' 
{\bf Phys. Rev. Lett., 80} (1998).
%
\bibitem{ref6} S. Ghosh, T. F. Rosenbaum, G. Aeppli and S. N. Coppersmith. ``{\it Entangled quantum state of magnetic dipoles}'' 
{\bf Nature, 452} (2003).
%
\bibitem{ref7} P. Ball. ``{\it The dawn of quantum biology}'' 
{\bf Nature, 474} (2011).
%
\bibitem{ref8} S. Bose, A. Mazumdar, G. W. Morley, H. Ulbricht, M. Toros, M. Paternostro, A. A. Geraci, P. F. Barker, M. S. Kim, and G. Milburn. ``{\it Spin Entanglement Witness for Quantum Gravity.,}'' 
{\bf Phys. Rev. Lett., 119} (2017).
%
\bibitem{ref9} M. Wie\'sniak, V. Vedral and C. Brukner. ``{\it Magnetic susceptibility as a macroscopic entanglement 
witness,}'' 
{\bf New J. Phys., 7} (2005) 258.
%
\bibitem{ref10} O. M. Del Cima, D. H. T. Franco and S. L. L. Silva. ``{\it Quantum entanglement in trimer spin-1/2 Heisenberg chains with antiferromagnetic coupling,}'' 
{\bf Quantum Stud.: Math. Found., 3} (2015) DOI 10.1007/s40590-015-0059-1.
%
\bibitem{refLi} Y. Li and G. Zhu. ``{\it Concurrence vectors for entanglement of high-dimensional systems,}'' 
{\bf Front. Phys. China., 3} (2015).
%
\bibitem{refOst} A. Osterloh. ``{\it SL-invariant entanglement measures in higher dimension: the case of spin $1$ and $3/2$,}'' 
{\bf J. Phys. A, 48} (2017) DOI 10.1007/s40509-017-0149-3.
%
\bibitem{refBah} H. Bahmani, G. Najarbashi and A. Tavana. ``{\it Generalized concurrence and quantum phase transition in spin-$1$ Heisenberg model,}'' 
{\bf Phys. Scr., 95} (2020).
%
\bibitem{refSch} A. Scheie, P. Laurell, A. M. Samarakoon, B. Lake, S. E. Nagler, G. E. Granroth, S. Okamoto, G. Alvarez and D. A. Tennat. ``{\it Witnessing entanglement in quantum magnets using neutron scattering,}'' 
{\bf arXiv:210.08376v1} (2021).
%
\bibitem{refSau} S. L. L. Silva. ``{\it Thermal entanglement in $2\otimes3$ Heisenberg chains via distance between states,}'' 
{\bf arXiv:2103.02019v1} (2021).
%
\bibitem{refDahl} G. Dahl, J. M. Leinaas, J. Myrhein and E. Ovrum. ``{\it A tensor product matrix approximation problem in quantum physics,}'' 
{\bf Linear Algebra and its aplications, 420} (2007).
%
\bibitem{ref11} S. L. L. Silva. ``{\it Entanglement of spin-$1/2$ Heisenberg antiferromagnetic quantum spin chains,}'' 
{\bf Quantum Stud.: Math. Found. 5} (2017) DOI 10.1007/s40509-017-0149-3.
%
\bibitem{ref12} O. M. Del Cima, D. H. T. Franco and M M. Silva. ``{\it Magnetic shielding of quantum entanglement states,}'' 
{\bf Quantum Stud.: Math. Found., 6} (2019)DOI 10.1007/s40509-018-0172-z.
%
\bibitem{ref13} R. A. Bertlmann, H.Narnhofer and W. Thirring. ``{\it Geometric picture of entanglement and Bell inequalities,}'' 
{\bf Phys. Rev. A, 66} (2002). 
%
\bibitem{Wootters} K. M. O'Connor and W. K. Wootters. ``{\it Entanglement rings,}'' 
{\bf Phys. Rev. A, 63} (2001).
%

\end{thebibliography}

\end{document}